\documentclass{ifacconf}

\usepackage{graphics} 
\usepackage{epsfig} 
\usepackage{amsmath} 
\usepackage{amssymb}  
\usepackage{algpseudocode}
\usepackage{algorithm}
\usepackage{epstopdf}

\usepackage{verbatim}
\usepackage{color}
\usepackage{natbib}        
\usepackage{setspace}

\usepackage{flexisym}
\usepackage{breqn}

\begin{document}
\setstretch{1.07}
\begin{frontmatter}
\title{Stability and performance in MPC using a finite-tail cost\thanksref{footnoteinfo}} 

\thanks[footnoteinfo]{The authors thank the German Research Foundation (DFG) for support of this work within Soft Tissue Robotics (GRK 2198/1).\\
\textcopyright 2021 the authors. This work has been accepted to IFAC for publication under a Creative Commons Licence CC-BY-NC-ND. }

\author[First,Second]{Johannes K\"ohler}
\author[First]{Frank Allg\"ower}

\address[First]{Institute for Systems Theory and Automatic Control, University of Stuttgart, Germany (email:$\{$jokoehler,allgower$\}$@ist.uni-stuttgart.de).}
\address[Second]{Institute for Dynamic Systems and Control, ETH Zürich, Switzerland. (email:jkoehle@ethz.ch).}

\begin{abstract}
In this paper, we provide a stability and performance analysis of model predictive control (MPC) schemes based on finite-tail  costs.
We study the MPC formulation originally proposed by~\cite{magni2001stabilizing} wherein the standard terminal penalty is replaced by a finite-horizon cost of some stabilizing control law. 
In order to analyse the closed loop, we leverage the more recent technical machinery developed for MPC without terminal ingredients.   
For a specified set of initial conditions, we obtain sufficient conditions for stability and a performance bound in dependence of the prediction horizon and the extended horizon used for the terminal penalty.
The main practical benefit of the considered finite-tail cost MPC formulation is the simpler offline design in combination with typically significantly less restrictive bounds on the prediction horizon to ensure stability. 
We demonstrate the benefits of the considered MPC formulation using the classical example of a four tank system. 
\end{abstract} 
\end{frontmatter} 
\section{Introduction}
Model predictive control (MPC)  is an optimization-based control technique that is applicable to general nonlinear constrained systems~\citep{rawlings2017model,grune2017nonlinear}. 
The standard method to design a stabilizing MPC scheme includes the offline design of terminal ingredients, i.e., a local control Lyapunov function (CLF) and a terminal set ~\citep{chen1998quasi,mayne2000constrained}. 
However, terminal ingredients are often not employed in practical implementation due to challenges in the offline design, compare the discussion by~\cite{mayne2013apologia}. 
Specifically, while there exist simple to apply design methods based on the linearization around the steady-state~\citep{chen1998quasi,rawlings2017model}, the resulting terminal ingredients are often conservative. 
Furthermore, if a more general setup is considered, the design of the terminal ingredients becomes increasingly challenging, compare, e.g., online changing setpoints~\citep{limon2018nonlinear}, dynamic reference trajectories~\citep{faulwasser2012optimization,koehler2020nonlinearTAC}, or online model adaptation~\citep[Criterion 1--2]{adetola2011robust}.  
Thus, it is desirable to develop MPC schemes with guaranteed stability that relax the offline design requirements to allow for an easier implementation.

\subsubsection*{Related work}
\cite{alamir1995stability} showed that there exists some large enough prediction horizon, such that an MPC scheme without any terminal ingredients ensures asymptotic stability, compare also the work by~\cite{primbs2000feasibility} and \cite{jadbabaie2005stability}. 
Computable lower bounds on a sufficiently long prediction horizon using \textit{exponential cost controllability} have been derived by~\cite{tuna2006shorter,grune2008infinite} and later improved by~\cite{grune2009analysis,grune2010analysis} using a linear programming (LP) analysis, compare also the continuous-time results by~\cite{reble2012unconstrained}. 
Results for MPC without terminal ingredients using positive semi-definite stage costs are obtained by~\cite{grimm2005model} using the notion of cost detectability, compare also the less conservative results by~\cite{Koehler2020Regulation} using the notion of cost observability.

Stability based on infinite-horizon tail costs has been shown by~\cite{de1998stabilizing}.
 \cite{magni2001stabilizing} derived stability conditions for finite-tail cost MPC. 
The corresponding finite-horizon tail cost can be viewed as a relaxed CLF.  
By using a relaxed CLF as a terminal cost, stability can be established using a smaller prediction horizon, compare the analysis in~\cite[A3]{tuna2006shorter}, \cite[Ass.~5]{grimm2005model}, and \cite[Prop.~2]{reble2012improved}. 
Furthermore, by choosing the stage cost as a relaxed CLF with a non-uniform weighting, stability can be ensured for arbitrary short horizons~\citep{alamir2018stability}. 

\cite{boccia2014stability} considered a relaxed \textit{local} exponential cost controllability and derived a sufficiently long prediction horizon for a specified region of attraction, compare also the continuous-time results by~\cite{esterhuizen2020recursive}.
The corresponding analysis was further improved by~\cite{kohlernonlinear19} to derive less conservative bounds. 
These derivations all use sublevel set arguments for the value function, similar to the implicit terminal set constraint proposed by~\cite{limon2006stability}.

Overall, the resulting bounds on the prediction horizon for MPC without terminal ingredients are often too conservative for practical applications. 
Thus, providing easy to implement MPC schemes with guaranteed stability remains an open problem.

\subsubsection*{Contribution} 
We analyse the closed-loop stability and performance of the finite-tail cost MPC proposed by~\cite{magni2001stabilizing} using modern tools developed for MPC without terminal ingredients. 
In particular, we show that the finite-tail cost constitutes a relaxed CLF and use (local) exponential cost controllability to prove asymptotic stability and derive a suboptimality bound. 
Given a (local) stabilizing control law, an extended horizon for the finite-tail cost, and  a suitably chosen region of attraction,  we provide a lower bound on the prediction horizon that ensures closed-loop stability and a desired suboptimality bound for the closed-loop performance (relative to the infinite-horizon optimal performance). 

From a practical point of view, the considered MPC formulation can be viewed as an intermediate design between MPC schemes with and without terminal ingredients. 
The provided analysis reveals that the explicit inclusion of a (locally) stabilizing control law in the MPC formulation can significantly improve the resulting bounds. 
Thus, we can ensure closed-loop stability with a shorter prediction horizon and reduce the online computational demand. 
Compared to the design of a (local) CLF, the considered formulation only requires the offline computation of a (locally) stabilizing control law, which is often already available, especially in case of open-loop stable systems. 
The benefits and applicability of the proposed MPC formulation are illustrated using the classical four tank system. 


\subsubsection*{Notation}
The interior of a set $\mathbb{Z}$ is denoted by $\text{int}(\mathbb{Z})$.
The set of non-negative real numbers and integers are denoted by $\mathbb{R}_{\geq 0}$ and $\mathbb{I}_{\geq 0}$, respectively. 
The set of integers in the interval $[a,b]\subseteq\mathbb{R}$ are denoted by $\mathbb{I}_{[a,b]}$. 
By $\mathcal{K}_\infty$, we denote the set of continuous functions $\alpha:\mathbb{R}_{\geq 0}\rightarrow\mathbb{R}_{\geq 0}$ that are strictly increasing, unbounded, and satisfy $\alpha(0)=0$.
The quadratic norm with respect to a positive definite matrix $Q=Q^\top$ is denoted by $\|x\|_Q^2=x^\top Q x$.

\section{Finite-tail cost MPC}
\label{sec:MPC}
In this section, we describe the problem setup (Sec.~\ref{sec:MPC_1}), the stabilizing control law (Sec.~\ref{sec:MPC_2}), and the considered MPC formulation (Sec.~\ref{sec:MPC_3}). 
\subsection{Setup}
\label{sec:MPC_1}
We consider the following nonlinear discrete-time system
\begin{align*}
x(t+1)&=f(x(t),u(t)),\quad x(0)=x_0,
\end{align*}
with the state $x\in\mathbb{X}=\mathbb{R}^n$, the control input $u(t)\in\mathbb{U}\subseteq\mathbb{R}^m$, the dynamics $f:\mathbb{X}\times\mathbb{U}\rightarrow\mathbb{X}$, the initial condition $x_0\in\mathbb{X}$, and the time step $t\in\mathbb{I}_{\geq 0}$.
We impose point-wise in time constraints on the state and input
\begin{align*}
(x(t),u(t))\in \mathbb{Z}\subseteq\mathbb{X}\times\mathbb{U},\quad t\in\mathbb{I}_{\geq 0}.
\end{align*}
We consider a stage cost $\ell:\mathbb{X}\times\mathbb{U}\rightarrow\mathbb{R}_{\geq 0}$ and define $\ell_{\min}(x):=\inf_{u\in\mathbb{U}}\ell(x,u)$. 
The control goal is to minimize the closed-loop cost, stabilize the origin, and satisfy the constraints. 
We assume that $f(0,0)=0$, $0\in\text{int}(\mathbb{Z})$, $f,\ell$ are continuous, and $\mathbb{U}$ is compact.  
\begin{assum} (Tracking stage cost)
\label{ass:tracking_cost}
There exist functions $\underline{\alpha}_\ell, \overline{\alpha}_\ell\in\mathcal{K}_\infty$, such that $\underline{\alpha}_\ell(\|x\|)\leq \ell_{\min}(x)\leq \overline{\alpha}_\ell(\|x\|)$ for all $x\in\mathbb{X}$ and $\ell(0,0)=0$. 
\end{assum}
This assumption characterizes the fact that the stage cost is positive definite w.r.t. the origin and is typically ensured by choosing a quadratic stage cost.

\subsection{Stabilizing feedback}
\label{sec:MPC_2}
The considered MPC formulation uses a known continuous control law $\kappa:\mathbb{X}\rightarrow\mathbb{U}$ to define a finite-tail cost. 
The corresponding stage cost is denoted by $\ell_\kappa(x):=\ell(x,\kappa(x))$. 
Given some initial state $x$, we define the closed-loop state and input response under the feedback $\kappa$ by the continuous functions $\phi_{\mathrm{x}}:\mathbb{I}_{\geq 0}\times\mathbb{X}\rightarrow\mathbb{X}$ and $\phi_{\mathrm{u}}:\mathbb{I}_{\geq 0}\times\mathbb{X}\rightarrow\mathbb{U}$: 
\begin{align*}
\phi_{\mathrm{x}}(0,x):=&x,\quad 
\phi_{\mathrm{u}}(k,x):=\kappa(\phi_{\mathrm{x}}(k,x)),\quad k\in\mathbb{I}_{\geq 0},\\
\phi_{\mathrm{x}}(k+1,x):=&f(\phi_{\mathrm{x}}(k,x),\phi_{\mathrm{u}}(k,x)),\quad k\in\mathbb{I}_{\geq 0}.
\end{align*}
%
\begin{assum}(Locally stabilizing controller)
\label{ass:exp_feedback}
There exist constants $\rho\in[0,1)$, $C\geq 1$, $\epsilon>0$ such that for all $x\in\mathbb{X}$ satisfying $\ell_{\min}(x)\leq \epsilon$, we have
\begin{subequations}
\label{eq:exp_feedback}
\begin{align}
\label{eq:exp_feedback_cost}
&\ell_\kappa(\phi_{\mathrm{x}}(k,x))\leq C\rho^{k} \ell_{\min}(x),\quad k\in\mathbb{I}_{\geq 0},\\
\label{eq:exp_feedback_constraint}
&(\phi_{\mathrm{x}}(k,x),\phi_{\mathrm{u}}(k,x))\in\mathbb{Z},\quad k\in\mathbb{I}_{\geq 0}.
\end{align}
\end{subequations}
\end{assum}
In case a quadratic stage cost $\ell$ is used, Condition~\eqref{eq:exp_feedback_cost} holds if the controller $\kappa$ ensures (local) exponential stability, compare~\cite[Sec.~6.2]{grune2017nonlinear}. 
Condition~\eqref{eq:exp_feedback_constraint} follows from $0\in\text{int}(\mathbb{Z})$, $\kappa$ continuous, Condition~\eqref{eq:exp_feedback_cost} and Assumption~\ref{ass:tracking_cost} for $\epsilon>0$ small enough. 
Hence, a suitable controller $\kappa$ exists if the linearization at the origin is stabilizable, c.f.~\citep{chen1998quasi}.

\subsection{MPC formulation}
\label{sec:MPC_3}
Given some horizon length $M\in\mathbb{I}_{\geq 1}$, the finite-tail cost $V_{\mathrm{f},M}:\mathbb{X}\rightarrow\mathbb{R}_{\geq 0}$ is defined as
\begin{align*}
&V_{\mathrm{f},M}(x):=\\
&\begin{cases}
\sum_{k=0}^{M-1}\ell_\kappa(\phi_{\mathrm{x}}(k,x)) & \text{if } (\phi_{\mathrm{x}}(k,x),\phi_{\mathrm{u}}(k,x))\in\mathbb{Z}, k\in\mathbb{I}_{[0,M-1]} \\
\infty & \mathrm{otherwise} 
\end{cases}\nonumber
\end{align*}
Given a predicted state and input sequence $x(\cdot|t)\in\mathbb{X}^{N+1}$, $u(\cdot|t)\in\mathbb{U}^N$, the considered cost function is given by
\begin{align*}
\mathcal{J}_{N,M}(x(\cdot|t),u(\cdot|t)):=&\sum_{k=0}^{N-1}\ell(x(k|t),u(k|t))+V_{\mathrm{f},M}(x(N|t)),
\end{align*}
where $N\in\mathbb{I}_{\geq 1}$ is the prediction horizon and $M\in\mathbb{I}_{\geq 1}$ is the extended horizon used for the finite-tail cost. 
At time $t$, given a state $x(t)$, the MPC control law is determined based on the following optimization problem
\begin{align}
\label{eq:MPC}
\inf_{u(\cdot|t)}&\mathcal{J}_{N,M}(x(\cdot|t),u(\cdot|t))\\
\text{s.t. }&x(k+1|t)=f(x(k|t),u(k|t)),~ k\in\mathbb{I}_{[0,N]},\nonumber\\
&x(0|t)=x(t),~ 
(x(k|t),u(k|t))\in\mathbb{Z},~ k\in\mathbb{I}_{[0,N-1]}.\nonumber
\end{align}
A minimizer\footnote{%
Given feasibility, a minimizer exists due to continuity of $\ell,\kappa,f$ and $\mathbb{U}$ compact, cf.~\citep[Prop~2.4]{rawlings2017model}. 
We define $V_{N,M}(x(t))=\infty$ in case the optimization problem~\eqref{eq:MPC} is not feasible. 
} is denoted by $u^*(\cdot|t)$ with the corresponding state trajectory $x^*(\cdot|t)$ and the value function $V_{N,M}(x(t)):=\mathcal{J}_{N,M}(x^*(\cdot|t),u^*(\cdot|t))$.  
The closed-loop is given by applying the initial part of an optimal input sequence, i.e., $u(t)=u^*(0|t)$. 

Compared to an MPC scheme using standard terminal ingredients~\citep{chen1998quasi,mayne2000constrained}, the considered MPC formulation~\eqref{eq:MPC} drops the terminal set constraint (as also proposed by~\cite{limon2006stability}) and the standard terminal cost is replaced by a finite-tail cost $V_{\mathrm{f},M}$. 
Due to the specific structure of the finite-tail cost $V_{\mathrm{f},M}$, the considered MPC formulation can also be viewed as an MPC scheme without terminal ingredients~\citep{grune2009analysis} with a prediction horizon $N+M$, subject to the additional parametrization $u(k|t)=\kappa(x(k|t))$, $k\in\mathbb{I}_{[N,N+M-1]}$, similar to move-blocking~\citep{cagienard2007move}.

The only offline design required for the MPC implementation is the stabilizing feedback $\kappa$. 
The prediction horizon $N$ and the extended horizon $M$ are tuning variables that trade of stability, performance, and computational complexity, compare the discussion in Remark~\ref{rk:trade_off} below.

\section{Theoretical analysis}
\label{sec:theory}
In the following, we study the closed-loop properties of the finite-tail cost MPC. 
First, Propositions~\ref{prop:gamma} and \ref{prop:CLF} derive bounds on the value function and the finite-tail cost.
Then, Theorem~\ref{thm:main} provides a closed-loop analysis. 
%
\begin{prop}
\label{prop:gamma}
Let Assumption~\ref{ass:exp_feedback} hold. 
For any $N\in\mathbb{I}_{\geq 0}$, $M\in\mathbb{I}_{\geq 0}$, for any $x\in\mathbb{X}$ with $\ell_{\min}(x)\leq \epsilon$, the optimization problem~\eqref{eq:MPC} is feasible and the value function satisfies
\begin{align}
\label{eq:gamma_bound}
V_{N,M}(x)\leq \gamma_{N+M}\ell_{\min}(x),~ \gamma_{k}:=\frac{1-\rho^{k}}{1-\rho}C,~k\in\mathbb{I}_{\geq  0}.
\end{align}
\end{prop}
\begin{pf}
Condition~\eqref{eq:exp_feedback_constraint} in Assumption~\ref{ass:exp_feedback} ensures that $u(k|t)=\kappa(x(k|t))$ is locally a feasible candidate solution to~\eqref{eq:MPC} ensuring $V_{N,M}(x)\leq V_{\mathrm{f},N+M}(x)$. 
Condition~\eqref{eq:exp_feedback_cost} and the geometric series yield the expression for $\gamma_{k}$ in~\eqref{eq:gamma_bound}. 
$\hfill\square$
\end{pf}
Condition~\eqref{eq:gamma_bound} corresponds to the (local) exponential cost controllability condition that is used in the analysis of MPC without terminal ingredients~\citep{boccia2014stability}. 
%
%
\begin{prop}
\label{prop:CLF}
Let Assumption~\ref{ass:exp_feedback} hold. 
For any $M\in\mathbb{I}_{\geq 1}$ and any $x\in\mathbb{X}$ satisfying $\ell_{\min}(x)\leq \epsilon$, we have
\begin{subequations}
\label{eq:relaxed_CLF}
\begin{align}
\label{eq:relaxed_CLF_1}
\ell_{\min}(x)\leq V_{\mathrm{f},M}(x)\leq& \gamma_M\ell_{\min}(x),\\
\label{eq:relaxed_CLF_2}
V_{\mathrm{f},M+1}(x)-V_{\mathrm{f},M}(x)\leq& C\rho^M\ell_{\min}(x).
\end{align}
If additionally $V_{\mathrm{f},M}(x)\leq \epsilon$ holds, then we have
\begin{align}
\label{eq:relaxed_CLF_4}
V_{\mathrm{f},M+1}(x)\leq&(1+c_M)V_{\mathrm{f},M}(x),~c_M:=\frac{C\rho^M(1-\rho)}{1-\rho^M}.
\end{align}
\end{subequations}
\end{prop}
\begin{pf}
The lower and upper bound in~\eqref{eq:relaxed_CLF_1} follow by the definition of $\ell_{\min}$ and the proof in Proposition~\ref{prop:gamma}. 
Inequality~\eqref{eq:relaxed_CLF_2} follows using
\begin{align*}
V_{\mathrm{f},M+1}(x)-V_{\mathrm{f},M}(x)=\ell_{\kappa}(\phi_{\mathrm{x}}(M,x))\stackrel{\eqref{eq:exp_feedback_cost}}{\leq} C\rho^M\ell_{\min}(x).
\end{align*}
Note that given $\ell_\kappa(x)\leq V_{\mathrm{f},M}(x)$, Inequality~\eqref{eq:relaxed_CLF_4} directly follows from Inequality~\eqref{eq:relaxed_CLF_2} for $c_M=C\rho^M$, which is, however, conservative. 
In the following, we use an LP analysis similar to the one proposed by~\cite{grune2009analysis,grune2010analysis} to prove Inequality~\eqref{eq:relaxed_CLF_4}  with $c_M=C\rho^M\dfrac{1-\rho}{1-\rho^M}\leq C\rho^M$. 
Denote $\ell_k=\ell_\kappa(\phi_{\mathrm{x}}(k,x))$, $k\in\mathbb{I}_{[0,M]}$. 
Condition~\eqref{eq:relaxed_CLF_4} is equivalent to $\ell_M\leq c_M\sum_{k=0}^{M-1} \ell_k$. 
Assumption~\ref{ass:exp_feedback} and $\ell_j\leq V_{\mathrm{f},M}(x)\leq \epsilon$, $j\in\mathbb{I}_{[0,M-1]}$ ensure
\begin{align*}
\ell_k\leq C\rho^{k-j}\ell_j,~j\in\mathbb{I}_{[0,M-1]},~k\in\mathbb{I}_{[j+1,M]}. 
\end{align*}
We normalize the stage cost using $\tilde{\ell}_k:=\ell_k/V_{\mathrm{f},M}(x)$, $k\in\mathbb{I}_{[0,M]}$, where we assume w.l.o.g. $V_{\mathrm{f},M}(x)>0$. 
Thus, a constant $c_M$ satisfying~\eqref{eq:relaxed_CLF_4} can be computed using the following LP:
\begin{subequations}
\label{eq:LP}
\begin{align}
c_M:=&\max_{\tilde{\ell}_k} \tilde{\ell}_M\\
\text{s.t. }
&\sum_{k=0}^{M-1}\tilde\ell_k=1,\quad 
\tilde\ell_M\leq C\rho^{M-k}\tilde\ell_k,~k\in\mathbb{I}_{[0,M-1]}.
\end{align}
\end{subequations}
Using a standard argument of contradiction, the constraints are all active. 
Thus, the maximum can be analytically computed, yielding $c_M:=\dfrac{C\rho^M(1-\rho)}{1-\rho^M}$. $\hfill\square$
\end{pf}
For $M>\underline{M}:=\log(C)/\log(1/\rho)$, we have $\alpha_M:=1-C\rho^M>0$ and thus Inequality~\eqref{eq:relaxed_CLF_2} ensures that $V_{\mathrm{f},M}$ is (locally) a relaxed CLF, i.e., $V_{\mathrm{f},M}(f(x,\kappa(x)))\leq V_{\mathrm{f},M}(x)-\alpha_M\ell_\kappa(x)$ holds. 
For $M\rightarrow\infty$ we recover the infinite-tail cost proposed by~\cite{de1998stabilizing} which satisfies the standard CLF condition employed in the literature~\citep{rawlings2017model}. 
%
%
\begin{thm}
\label{thm:main}
Let Assumptions~\ref{ass:tracking_cost}--\ref{ass:exp_feedback} hold. 
Then, for any $\overline{V}>0$, $M\in\mathbb{I}_{\geq 1}$, there exists a constant $N_{\overline{V},M}\geq 0$ such that for all $N> N_{\overline{V},M}$ and any initial condition $x_0\in\mathbb{X}_{\overline{V}}:=\{x\in\mathbb{X} \mid V_{N,M}(x)\leq \overline{V}\}$, Problem~\eqref{eq:MPC} is feasible for all $t\in\mathbb{I}_{\geq 0}$, the constraints are satisfied, and the origin is asymptotically stable for the resulting closed loop. 
Furthermore, there exist constants $\alpha_{N,M}\in(0,1]$, $\epsilon_{N,M}\in(0,1]$ such that the following performance bound holds for the resulting closed loop:
\begin{align}
\label{eq:performance_finite_tail}
\sum_{k=0}^\infty \ell(x(k),u(k))\leq\dfrac{1}{\epsilon_{N,M}}V_{N,M}(x_0)
\leq \dfrac{1}{\alpha_{N,M}}V_{\infty,0}(x_0).
\end{align}
\end{thm}
\begin{pf}
The following proof uses the properties in Propositions~\ref{prop:gamma}-\ref{prop:CLF},  improved arguments regarding the characterization of the region of attraction that combine ideas from~\cite{kohlernonlinear19} and~\cite{limon2006stability}, and standard bounds for MPC without terminal ingredients based on exponential cost controllability~\citep{tuna2006shorter,grune2008infinite}. 
In order to simplify the following exposition, we use the constant bound $\gamma=\gamma_\infty=C/(1-\rho)$, which is valid for all horizons $N,M\in\mathbb{I}_{\geq 0}$.
The proof is split into three parts.  
Part I and II show that the value function $V_{N,M}$ satisfies the following bounds at time $t\in\mathbb{I}_{\geq 0}$, assuming $x(t)\in\mathbb{X}_{\overline{V}}$:
\begin{subequations}
\label{eq:UCON_finite_tail_1}
\begin{align}
\label{eq:UCON_finite_tail_1_1}
\ell_{\min}(x(t))\leq V_{N,M}(x(t))\leq &\gamma_{\overline{V}}\ell_{\min}(x(t)),\\
\label{eq:UCON_finite_tail_1_2}
V_{N,M}(x(t+1))-V_{N,M}(x(t))\leq &-\epsilon_{N,M}\ell(x(t),u(t)),
\end{align}
\end{subequations}
with later specified constants $\gamma_{\overline{V}}\geq \epsilon_{N,M}>0$. 
Part III establishes that $x(t)\in\mathbb{X}_{\overline{V}}$ holds recursively for all $t\in\mathbb{I}_{\geq 0}$, derives the performance bound~\eqref{eq:performance_finite_tail}, and establishes asymptotic stability.
 Abbreviate $\ell(k|t):=\ell(x^*(k|t),u^*(k|t))$ and $V(k|t):=V_{N-k,M}(x^*(k|t))$, $k\in\mathbb{I}_{[0,N]}$ with $u^*(N|t)=\kappa(x^*(N|t))$. \\
\textbf{Part I:} The lower bound in Inequality~\eqref{eq:UCON_finite_tail_1_1} follows directly from $\ell\geq 0$. 
Define $\gamma_{\overline{V}}:=\max\left\{\gamma,\frac{\overline{V}}{\epsilon}\right\}$. 
The upper bound in Inequality~\eqref{eq:UCON_finite_tail_1_1} follows from the local upper bound in Proposition~\ref{prop:gamma} and $V_{N,M}(x(t))\leq \overline{V}$ using a case distinction whether $\ell_{\min}(x(t))\leq \epsilon$, compare also~\citep{boccia2014stability,kohlernonlinear19}. \\
\textbf{Part II:} 
In the following, we show that there exists a point $k\in\mathbb{I}_{[0,N_0]}$, $N_0:=\left\lceil\gamma_{\overline{V}}-\gamma\right\rceil=\left\lceil\max\left\{0,\frac{\overline{V}-\gamma\epsilon}{\epsilon}\right\}\right\rceil\in\mathbb{I}_{\geq 0}$, such that $V(k|t)\leq \gamma\epsilon$. 
For contradiction assume $V(k|t)>\gamma\epsilon$, $k\in\mathbb{I}_{[0,N_0]}$. 
Using Proposition~\ref{prop:gamma} this implies $\ell(k|t)>\epsilon$, $k\in\mathbb{I}_{[0,N_0]}$ since $\ell(k|t)\leq \epsilon$ would imply $V(k|t)\leq \gamma\ell(k|t)\leq \gamma\epsilon$.  
For any $k'\in\mathbb{I}_{[0,N]}$, the principle of optimality ensures
\begin{align*}
V_{N,M}(x(t))=\sum_{k=0}^{k'-1}\ell(k|t)+V(k'|t).
\end{align*}
Thus, $\ell(k|t)>\epsilon$, $k\in\mathbb{I}_{[0,N_0]}$ and $V(0|t)\leq \overline{V}$ yields 
\begin{align*}
V(N_0|t)<\overline{V}-\epsilon N_0\leq \overline{V}-\epsilon(\gamma_{\overline{V}}-\gamma)\leq \gamma\epsilon,
\end{align*}
 which contradicts $V(k|t)>\gamma\epsilon$, $k\in\mathbb{I}_{[0,N_0]}$. 
Denote the smallest element $k\in\mathbb{I}_{[0,N]}$, which satisfies $V(k|t)\leq \gamma\epsilon$ by $k_x$. 
Thus, we have $V(k|t)\leq V(k_x|t)\leq \gamma\epsilon$ for $k\in\mathbb{I}_{[k_x,N]}$. 
Furthermore, $V(k|t)\leq \gamma\epsilon$, $k\in\mathbb{I}_{[k_x,N]}$ implies $V(k|t)\leq \gamma\ell(k|t)$, $k\in\mathbb{I}_{[k_x,N]}$ using Proposition~\ref{prop:gamma} and a case distinction whether $\ell(k|t)\leq \epsilon$.  
Furthermore, it holds that
\begin{align}
\label{eq:V_N_0_finite_tail}
V(k_x|t)\leq \gamma\min\{\ell(k_x|t),\epsilon\}\leq \gamma\min\{\ell(0|t),\epsilon\}, 
\end{align}
where the second inequality follows from the definition of $k_x$, i.e., $\ell(0|t)\leq \epsilon$ implies $k_x=0$. 
In addition, $V(k_x|t)\leq \overline{V}$ implies $V(k_x|t)\leq \underline{\gamma}\epsilon$, $\underline{\gamma}:=\min\{\gamma,\overline{V}/\epsilon\}$. 
In the following, we show that Condition~\eqref{eq:UCON_finite_tail_1_2} holds with some $\epsilon_{N,M}\in(0,1]$. 
Define $\rho_{\gamma}:=\dfrac{\gamma-1}{\gamma}\in[0,1)$. 
Given that $V(k|t)\leq \gamma\ell(k|t)$, $k\in\mathbb{I}_{[k_x,N]}$, we can use the bounds by~\cite{tuna2006shorter,grune2008infinite} for the remaining horizon of length $N-k_x\geq N-N_0$ to show
\begin{align}
\label{eq:V_exp}
V(N|t)\leq \rho_\gamma^{N-k_x} V(k_x|t)\stackrel{\eqref{eq:V_N_0_finite_tail}}{\leq} \rho_\gamma^{N-N_0}\min\{\gamma\ell(0|t),\underline{\gamma}\epsilon\}. 
\end{align}
For $N\geq N_1:=N_0+\left\lceil \dfrac{\max\{\log(\underline{\gamma}),0\}}{\log(\gamma)-\log(\gamma-1)}\right\rceil$ this implies $V(N|t)\leq \epsilon$. 
Finally, we bound the value function in the next time step by appending the local control law $\kappa$ as a feasible candidate solution, yielding
\begin{align*}
&V_{N,M}(x(t+1))-V_{N,M}(x(t))\\
\leq& -\ell(0|t)+V_{\mathrm{f},M+1}(x^*(N|t))-V_{\mathrm{f},M}(x^*(N|t))\nonumber\\
\stackrel{\eqref{eq:relaxed_CLF_4}}{\leq} &-\ell(0|t)+c_MV(N|t)\nonumber \\
\stackrel{\eqref{eq:V_exp}}{\leq }&-\ell(0|t)+c_M \rho_\gamma^{N-N_0}\gamma \ell(0|t)\nonumber\\
\stackrel{ \mathrm{Prop.}~\ref{prop:gamma}-\ref{prop:CLF}}{=}&- \underbrace{\left(1-C^2 \rho_\gamma^{N-N_0}\dfrac{\rho^M }{1-\rho^M} \right)}_{=:\epsilon_{N,M}}\ell(0|t).
\end{align*}
For $N>N_2:=N_0+\dfrac{\log(C^2\rho^M/(1-\rho^M))}{\log(\gamma)-\log(\gamma-1)}$, we have $\epsilon_{N,M}>0$. 
Thus, given any $M\in\mathbb{I}_{\geq 1}$, all the previous bounds hold for  
\begin{align*}
N>N_M:=N_0+\dfrac{\max\{\log(\underline{\gamma}),\log(C^2\rho^M/(1-\rho^M)),0\}}{\log(\gamma)-\log(\gamma-1)}.
\end{align*}
 \textbf{Part III:} Condition~\eqref{eq:UCON_finite_tail_1_2} with $\ell\geq 0$ and $\epsilon_{N,M}>0$, ensure that $V_{N,M}$ is non-increasing and thus $x(t)\in\mathbb{X}_{\overline{V}}$ holds for all $t\in\mathbb{I}_{\geq 0}$. 
Hence, the results in Part I and II hold for all $t\in\mathbb{I}_{\geq 0}$. 
Inequalities~\eqref{eq:UCON_finite_tail_1} and Assumption~\ref{ass:tracking_cost} ensure asymptotic stability of the origin $x=0$. 
The first inequality in the performance bound~\eqref{eq:performance_finite_tail} follows directly by using Inequality~\eqref{eq:UCON_finite_tail_1_2} in a telescopic sum. 
Given an initial condition $x_0\in\mathbb{X}$, define an infinite horizon optimal trajectory $x_\infty(k)$, $u_\infty(k)$, $k\in\mathbb{I}_{\geq 0}$ with $V_{\infty,0}(x_0)=\sum_{k=0}^\infty \ell(x_\infty(k),u_\infty(k))$. 
Note that $V_{\infty,0}(x_0)\geq V_{N,M}(x_0)$ would directly imply~\eqref{eq:performance_finite_tail} with $\alpha_{N,M}=\epsilon_{N,M}$ and thus we  consider w.l.o.g. $V_{\infty,0}(x_0)\leq V_{N,M}(x_0)\leq \overline{V}$. 
We note that the local upper bound on the value function from Proposition~\ref{prop:gamma} is equally applicable to the infinite-horizon value function, i.e., $\ell_{\min}(x_{\infty}(k))\leq \epsilon$ implies $V_{\infty,0}(x_{\infty}(k))\leq \gamma\ell_{\min}(x_{\infty}(k))$, $k\in\mathbb{I}_{\geq0}$.  
Thus, we can  use the same steps from Part~II to show that the infinite-horizon optimal trajectory satisfies
\begin{align*}
V_{\infty,0}(x_\infty(N))\leq &\rho_\gamma^{N-N_0}\min\{\underline{\gamma}\epsilon,V_{\infty,0}(x_0)\}.
\end{align*}
We have $\ell_{\min}(x_\infty(N))\leq V_{\infty,0}(x_\infty(N))\leq\epsilon$ using $N\geq N_1$ and thus Inequality~\eqref{eq:relaxed_CLF_1} yields $V_{\mathrm{f},M}(x_\infty(N))\leq \gamma \ell_{\min}(x_\infty(N))$. 
The initial part of the infinite-horizon optimal trajectory is a feasible candidate solution to the optimization problem~\eqref{eq:MPC}, implying
\begin{align*}
V_{N,M}(x_0)\leq& V_{\infty,0}(x_0)+V_{\mathrm{f},M}(x_\infty(N))\\
\leq& (1+\gamma \rho_\gamma^{N-N_0})V_{\infty,0}(x_0).
\end{align*}
Inequality~\eqref{eq:performance_finite_tail} follows with $\alpha_{N,M}:=\dfrac{\epsilon_{N,M}}{1+\gamma\rho_\gamma^{N-N_0}}\in(0,\epsilon_{N,M}]$. $\hfill\square$
\end{pf}
Although the underlying MPC formulation including $V_{\mathrm{f},M}$ is essentially equivalent to the MPC formulation by~\cite{magni2001stabilizing}, the proof deviates significantly. 
In particular, in~\cite[Eq.~(13)]{magni2001stabilizing} it is simply assumed that 
\begin{align*}
\ell_\kappa(\phi_{\mathrm{x}}(M,x^*(N|t)))< \ell(x(t),u^*(0|t)).
 \end{align*}
 In our analysis, on the contrary, we provide explicit computable conditions in terms of lower bounds on the prediction horizon $N_M$ that ensure the desired closed-loop properties. 

The constant $\alpha_{N,M}$ bounds the suboptimality of the closed loop performance compared to the infinite-horizon optimal cost $V_{\infty,0}(x)$. 
Similar estimates are frequently used in MPC without terminal ingredients~\citep{reble2012unconstrained}, however, corresponding bounds for MPC with terminal ingredients are often less intuitive, compare the results by~\cite{grune2008infinite}.  
\begin{rem} (Comparison to state of the art)
\label{rk:compare}
The derived theoretical result is interesting for multiple reasons. 
First, it generalizes and unifies the standard results for MPC without terminal ingredients (\cite{boccia2014stability,grune2008infinite,grune2009analysis,grune2010analysis}, $M=0$) and MPC with terminal ingredients (\cite{chen1998quasi,de1998stabilizing,limon2006stability}, $M=\infty$), by considering a relaxed CLF (cf. Prop~\ref{prop:CLF}). 
Compared to an MPC formulation with terminal ingredients, weaker conditions are imposed on the terminal cost and no explicit terminal set constraint is used. 
The price we have to pay for this relaxation is that a horizon $N>N_M$ is required.
However, compared to the standard bounds in MPC without terminal ingredients (cf.~\cite{boccia2014stability,grune2008infinite,tuna2006shorter,grune2009analysis,grune2010analysis}, $M=0$), the bounds derived in Theorem~\ref{thm:main} can be significantly less conservative, as explained in the following. 
To simplify the following discussion, suppose that Assumption~\ref{ass:exp_feedback} holds globally (as assumed in most of the literature). 
Then, Theorem~\ref{thm:main} ensures stability, if $ \dfrac{\rho^{M}\rho_\gamma^{N}}{1-\rho^M}$ is sufficiently small.
For comparison, the simple bounds used in MPC without terminal ingredients~\citep{grune2008infinite,boccia2014stability,tuna2006shorter} require $\rho_\gamma^{N}$ to be small enough. 
The most important difference is the fact that the extended horizon $M$ scales w.r.t. $\rho$  while the prediction $N$ scales with $\rho_\gamma$, which is larger since $1-\rho_\gamma=\dfrac{1-\rho}{C}\leq 1-\rho$. 
This can lead to a significant reduction in the overall needed horizon $N+M$ as demonstrated in Section~\ref{sec:num} with a numerical example.
Furthermore, we can also ensure (local) stability with an arbitrary short prediction horizon $N$ by choosing $M$ large enough.  
The derivation of the region of attraction in Theorem~\ref{thm:main} reduces the conservatism inherent in earlier arguments used by~\cite{boccia2014stability,esterhuizen2020recursive} by splitting the prediction horizon $N$ in two intervals, combining the ideas by~\cite{kohlernonlinear19} and \cite{limon2006stability}.  
\end{rem}

\begin{rem} (Stability, performance, computation complexity)
\label{rk:trade_off}
The choice of the prediction horizon $N$ and the extended horizon $M$ are tuning variables that impact the stability ($\epsilon_{N,M}$), performance ($\alpha_{N,M}$), and the computational demand. 
In particular, increasing either $N$ or $M$ increases $\epsilon_{N,M}$ and thus improves the nominal stability. 
On the other hand, increasing the region of attraction or improving the suboptimality estimate $\alpha_{N,M}$ above a certain threshold can only be achieved by increasing the prediction horizon $N$. 
In particular, for $N\rightarrow\infty$ we can choose $\overline{V}$ arbitrary large and we approach infinite-horizon performance ($\alpha_{N,M}=1$), which is not the case for a large extended horizon $M$. 
This is rather natural, since the feedback $\kappa$ only ensures \textit{local} stability and we did not assume that the controller is optimal in any sense. 
From a computational perspective, increasing $N$ or $M$ similarly increases the number of function evaluations involved in the cost function and constraints. 
However, if a condensed formulation is used, an increase in $M$ does not increase the number of decision variables. 
Thus, a large extended horizon $M$ with a small prediction horizon $N$ can result in drastically reduced computational requirements compared to an MPC scheme without terminal ingredients that may require a significantly larger prediction horizon $N$. 
\end{rem}

\section{Numerical example}
\label{sec:num}
The following example shows the general applicability of the finite-tail cost MPC and the benefits compared to an MPC formulation without any terminal ingredients. 
We consider the nonlinear four tank system taken from~\cite{raff2006nonlinear} with
\begin{align*}
A_1\dot{x}_1=&-a_1\sqrt{2gx_1}+a_3\sqrt{2gx_3}+b_1 u_1,\\
A_2\dot{x}_2=&-a_2\sqrt{2gx_2}+a_4\sqrt{2gx_4}+b_2 u_2,\\
A_3\dot{x}_3=&-a_3\sqrt{2gx_3}+(1-b_2)u_2,\\
A_4\dot{x}_4=&-a_4\sqrt{2gx_4}+(1-b_1)u_1,
\end{align*}
where $x_i$ denotes the water level of tank $i$, $u_i$ corresponds to the water inflow, and the parameters $a_i,A_i,b_i,g$ are positive constants. 
The discrete-time system is defined using an Euler discretization and a sampling time of $T_{\mathrm{s}}=3s$.   
We consider the problem of stabilizing a setpoint $(x_{\mathrm{s}},u_{\mathrm{s}})$ with the quadratic stage cost $\ell(x,u)=\|x-x_{\mathrm{s}}\|_Q^2+\|u-u_{\mathrm{s}}\|_R^2$, $Q=\text{diag}(1,1,10^{-2}, 10^{-2})$, $R=\text{diag}(10^{-4},10^{-4})$. 
This problem is equivalent to the problem of stabilizing the origin (Ass.~\ref{ass:tracking_cost}), using a simple coordinate shift. 
The system constants, the setpoint, and the box constraints $\mathbb{Z}$ are chosen as in~\cite{raff2006nonlinear}.

In the following, the local stability properties of $\kappa$ are verified numerically, analogous to the numerical computation of the terminal set size in~\citep[Remark~3.1]{chen1998quasi}. 
Assumption~\ref{ass:exp_feedback} can be satisfied by choosing $\kappa$ as the LQR based on the linearization around the setpoint, resulting in $\rho=0.93$, $C=6.9$, $\epsilon=0.08$ and Inequality~\eqref{eq:gamma_bound} holds with $\gamma_{k}\leq74<98.3=C/(1-\rho)$ for $k\in\mathbb{I}_{[0,30]}$.

The system is open-loop stable with a rather slow system dynamics and thus might seem simple to stabilize using MPC. 
However, this system is in fact a standard example to demonstrate that an MPC scheme without any terminal ingredients is in general not stabilizing, unless the prediction horizon is chosen large enough, compare~\cite{raff2006nonlinear}.
In fact, using arguments similar to Theorem~\ref{thm:main} to guarantee local stability of the MPC without terminal ingredients requires $N> 2\frac{\log( \gamma)}{\log(\gamma)-\log(\gamma-1)}> 600$, which is quite conservative. 
\cite{grune2010analysis} reduced the conservatism of these bounds by using an LP analysis, an additional scalig $\omega>1$ for the stage cost, and implementing the first $K$ steps of the optimal open-loop input sequence. 
However, even with a weighting of $\omega=5\cdot 10^2$ and a multi-step implementation of $K=20$, the results in~\citep[Thm.~5.4]{grune2010analysis} still requires a prediction horizon of $N\geq 132$ to ensure stability. 
Thus, for the present example, an MPC scheme without any terminal ingredients requires a large prediction horizon $N$ to ensure stability which results in a large computational demand. 

We consider the finite-horizon cost with an extended horizon of $M=25$ and a prediction horizon of $N=5$. 
Inequality~\eqref{eq:relaxed_CLF_4} holds with $c_M=0.013$, which is significantly smaller than the bound $c_M=C\rho^M\frac{1-\rho}{1-\rho^M}\approx 0.09$ resulting from Proposition~\ref{prop:CLF}. 
Given these constants, Theorem~\ref{thm:main} ensures asymptotic stability for all initial condition $x_0\in\mathbb{X}_{\overline{V}}$ with $\overline{V}>\epsilon$. 
Thus, the proposed MPC formulation ensures stability in a reasonable region of attraction with a combined horizon of $N+M=30$. 
By considering the linearization, we find that even with $M=8$ and $N=1$ the finite-tail cost MPC yields a stable closed loop, however, with a potentially small region of attraction.

\begin{figure}[hbtp]
\begin{center}
\includegraphics[width=0.4\textwidth]{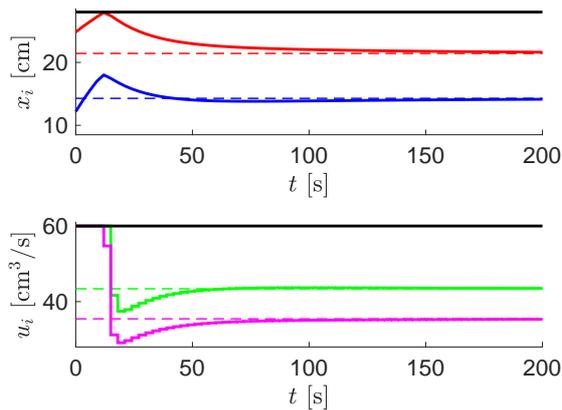}
\end{center}
\caption{Exemplary closed-loop trajectories. Top: $x_3$ (blue), $x_4$ (red) with steady-state $x_{\mathrm{s}}$ (dashed), and state constraint (black). 
Bottom: $u_1$ (green), $u_2$ (magenta), stationary input $u_{\mathrm{s}}$ (dashed), and input constraint (black).}
\label{fig:M}
\end{figure}

An exemplary closed-loop trajectory for $N=5$, $M=25$, and  $x_0=x_{\mathrm{s}}-(4,4,4,-2.5)$ can be seen in Figure~\ref{fig:M}. 
The MPC efficiently handles the state and input constraints by using the maximal control input to ensure fast convergence before reducing the control input to ensure satisfaction of the state constraint.
This example demonstrated the benefits of the finite-tail cost MPC, which requires little offline design (only a locally stabilizing feedback $\kappa$) and ensures asymptotic stability with a relatively short horizon compared to MPC without terminal ingredients.

\section{Conclusion}
\label{sec:sum}
We provided a stability and performance analysis of MPC based on finite-tail cost. 
The proposed MPC formulation provides a nice middle ground between MPC formulations with and without terminal ingredients. 
Due to the simple design and the typically less restrictive stability condition, this MPC formulation may be especially useful whenever the design of suitable terminal ingredients is non-trivial, such as control problems involving dynamic operation, model adaptation, or data-driven input-output models. 
 \section*{ACKNOWLEDGEMENTS}
We would like to thank our colleagues D. Gramlich and J. Berberich for helpful comments on the manuscript. 

\setstretch{1}
\bibliography{bibfile}  
\end{document}